\documentclass[aps,prl,twocolumn,superscriptaddress]{revtex4}

\usepackage{graphicx}
\usepackage{amsmath}
\usepackage{mathtools}
\usepackage{amssymb}
\usepackage{subfigure}

\usepackage[usenames]{color}

\begin{document}

\title{Mesoscopic structures and the Laplacian spectra of random geometric graphs}

\author{Amy Nyberg}
\affiliation{Department of Physics, University of Houston, Houston, Texas 77204-5005, USA}
\affiliation{Texas Center for Superconductivity, University of Houston, Houston, Texas 77204-5002, USA}
\author{Thilo Gross}
\affiliation{Department of Engineering Mathematics, University of Bristol, Merchant Venturers Building, Woodland Road, Bristol BS81UB, UK}
\author{Kevin E. Bassler}
\affiliation{Department of Physics, University of Houston, Houston, Texas 77204-5005, USA}
\affiliation{Texas Center for Superconductivity, University of Houston, Houston, Texas 77204-5002, USA}
\affiliation{Max Planck Institute for the Physics of Complex Systems, 
N\"{o}thnitzer Stra{\ss}e 38, Dresden D-01187, Germany}
\date{\today}

\begin{abstract}
We investigate the Laplacian spectra of random geometric graphs (RGGs). 
The spectra are found to consist of both a discrete and a continuous part.  
The discrete part is a collection of 
Dirac delta peaks at integer values roughly centered around the mean degree. 
The peaks are mainly due to the existence of mesoscopic structures that occur far more
abundantly in RGGs than in non-spatial networks.  
The probability of certain mesoscopic structures is analytically calculated 
for one-dimensional RGGs and they are shown to
produce integer-valued eigenvalues that comprise a significant fraction of the spectrum,
even in the large network limit.
A phenomenon reminiscent of Bose-Einstein condensation 
in the appearance of zero eigenvalues is also found.
\end{abstract}
\maketitle

Over the past two decades there has been considerable progress in the development of parameters and measurements to characterize complex networks. This has resulted in a rich description of both the microscopic 
and macroscopic properties of networks~\cite{albar, newmanrev, boccaletti}. However, until recently the intermediate, 
or mesoscale, level has not received the same degree of attention~\cite{chaosrev,ic}.  
The mesoscale level, though, is particularly interesting because
it is there that one can begin to understand how a network's modular 
structure affects its dynamics.   
From studies of the relationship between graph spectra and the 
structure and dynamics of networks~\cite{chung, beyond, pecora, van}  
it is known that certain mesoscale structures, namely symmetric (or quasi-symmetric) 
motifs reveal themselves in the spectrum~\cite{redundancy}. 
Symmetric motifs are of particular interest~\cite{symmetry}
because their spectral properties 
imply that the presence of a single such motif in a given network can have 
distinct, well-defined consequences for system-level processes such as diffusion, 
synchronization, or more complex dynamics~\cite{arenas, rgg_sync, lars, Ly}.
While symmetric structures are relatively rare in random non-spatial graphs, 
here we show that they occur abundantly in random geometric graphs (RGGs). 
RGGs result from randomly placing $N$ vertices in space and connecting those that are close
 and differ from other random graph models because of the metric that defines 
a distance between vertices~\cite{rgg_dall, penrose}.  
They are commonly used for modeling spatially 
embedded systems~\cite{spatial1,spatial2} such as wireless networks~\cite{wireless}, 
transportation and power grids~\cite{grids}, neural networks~\cite{toro1,toro2}, 
and certain biological 
processes~\cite{bio1,bio2}.  
Here, we show that the ensemble-averaged spectra of the graph Laplacian matrices of RGGs
indicate that mesoscopic symmetric structures occur abundantly in these graphs.

\begin{figure}
  \centering
  \includegraphics[width=.2\textwidth]{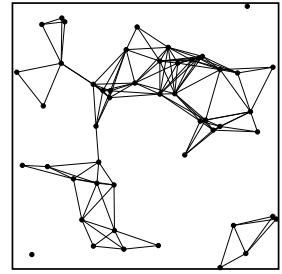}
  \caption{A RGG embedded in 2d on a square.} 
  \label{fig:2dpic}
\end{figure}
An example of a 2d RGG embedded on a square is shown in Fig.~\ref{fig:2dpic}.
Though to make progress analytically, we focus mostly on the Laplacian spectrum of 1d RGGs 
embedded on the unit circle, that is, 
in the domain [0,1] with periodic boundary conditions. 
Within this domain, the vertices are distributed randomly with 
uniform probability and two vertices are connected 
when the Euclidian distance between them is smaller than a threshold distance $r$. 
The discretized Laplace operator on the graph is given by the Laplacian matrix $\mathcal{L}$,
which has elements
$\mathcal{L}_{ij}=k_i \delta_{ij} - a_{ij}$, where $k_i$ is the degree 
of vertex $i$ and $a_{ij}$ is an element of the adjacency matrix, 
i.e., $a_{ij}=1$ if vertices $i$ and $j$ are connected and 0 otherwise. 
For this one-dimensional case, we analytically calculate the proportion 
of eigenvalues due to symmetry in the extensive and intensive scaling limits of large graphs.
Additionally, 
investigating the occurrence of the particular 
eigenvalue $\lambda=0$, we identify a phenomenon that is 
reminiscent of Bose-Einstein condensation. 

\begin{figure}
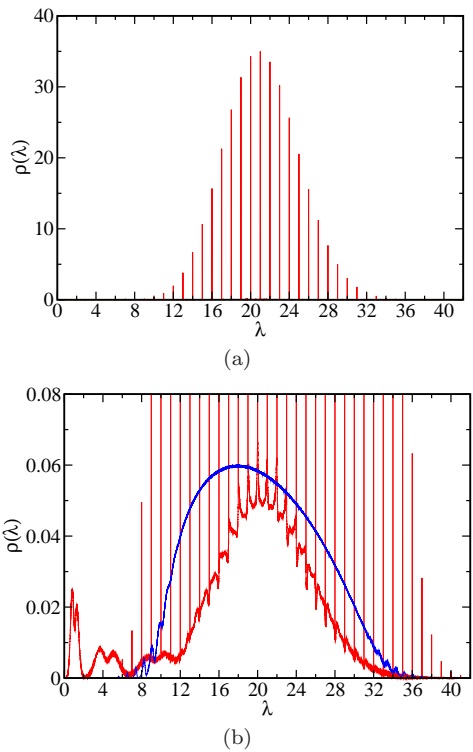

\begin{center}
\subfigure[]{\label{whole-b}\includegraphics[scale=.25]{zoomout.eps}} 
\subfigure[]{\label{whole-a}\includegraphics[trim = 0pt 0pt 0pt 0pt, clip=true, scale=.25]{rggercompare.eps}}
\caption{(Color online) The ensemble-averaged Laplacian spectrum for 1d RGGs on the unit circle (red),
with $N=100$ and $r=0.1$. 
(a) Top figure shows the
discrete, Dirac delta peaks that exist at integer eigenvalues. The envelope of these
peaks is centered about $\langle k \rangle +1$ where $\langle k \rangle$ = 20 is the
 mean degree of the network.
(b) The bottom figure is a zoom-in of the top figure.
Between the discrete peaks, the spectrum has a continuous part. 
The ensemble-averaged Laplacian spectum 
for Erd\H{o}s-R\'{e}nyi graphs with the same $N$ and $\langle k \rangle$ 
is shown for comparison (blue). 
These results are obtained from numerical diagonalization 
of $10^6$ realizations. A bin size of $\Delta \lambda = 0.001$ was used
to construct the histogram.  
}
	\label{whole}
\end{center}
\end{figure}

As can be seen in the example shown in Fig.~\ref{whole}, the spectra consist of discrete, 
Dirac delta peaks at integer eigenvalues
and a broad distribution of eigenvalues between the integers. 
The part of the spectra  between the integers becomes continuous
in the large network limit. We refer to the Dirac delta peaks as the discrete
part of the spectrum and the remainder as the continuous part.
Here, we focus on the discrete part.
As this figure suggests, and as we will analytically prove for 1d RGGs,
the discrete eigenvalues comprise a finite fraction of the spectra.
Note that the height of the Dirac delta peaks, but not the continuous part of the
spectra, depend on the bin size used.
The relative height of the peaks and, thus, the shape of their envelope is, however,
independent of bin size.  
By contrast, similar peaks are not visible in the ensemble-averaged Laplacian spectra of 
non-spatial random graph models such as the Erd\H{o}s-R\'enyi graphs, as the example shown
in Fig.~\ref{whole} indicates. 
\begin{figure}
\centering
\subfigure[]{\label{orbits-a}\includegraphics[scale=.25]{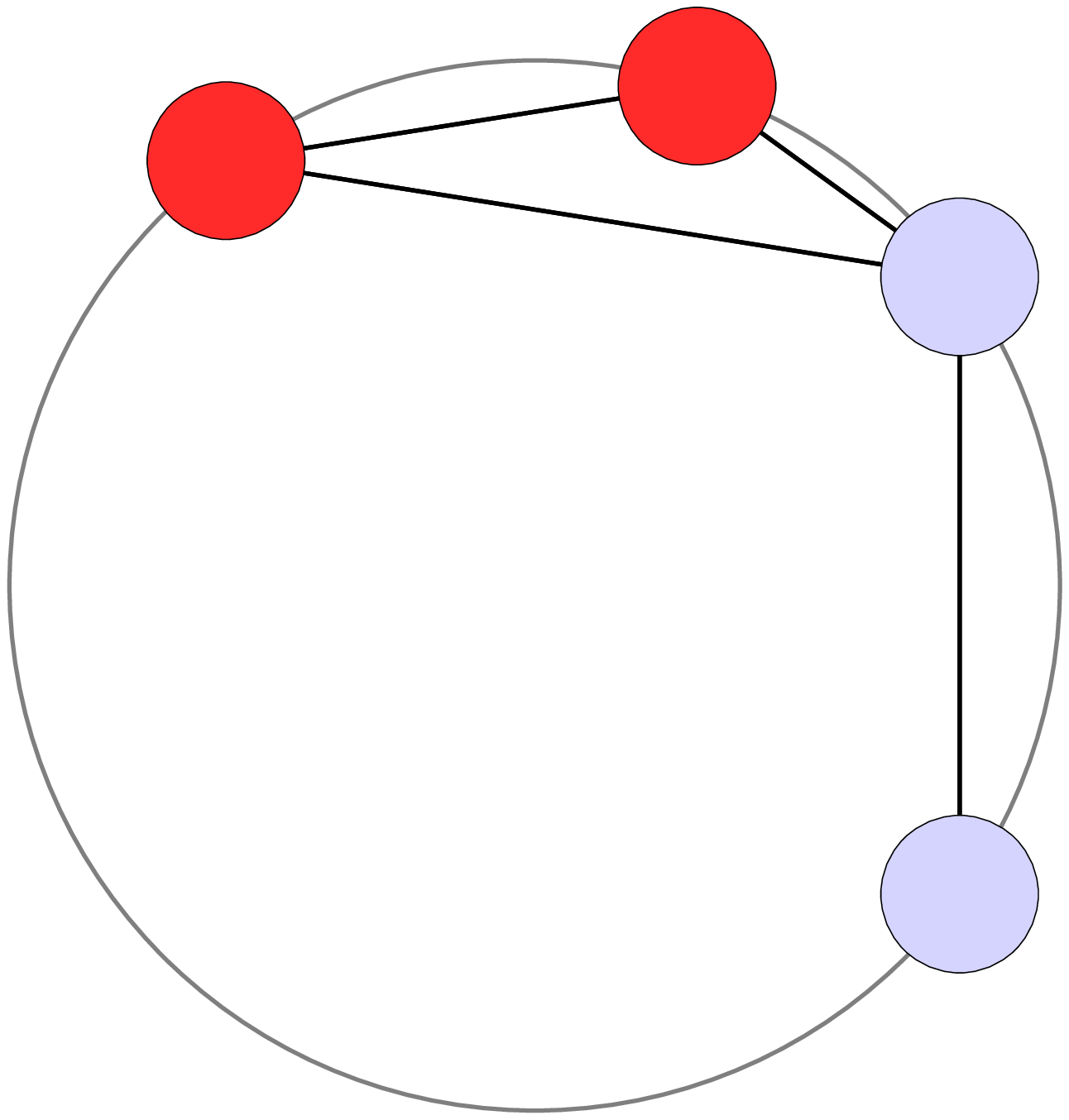}} \qquad
\subfigure[]{\label{orbits-b}\includegraphics[scale=.25]{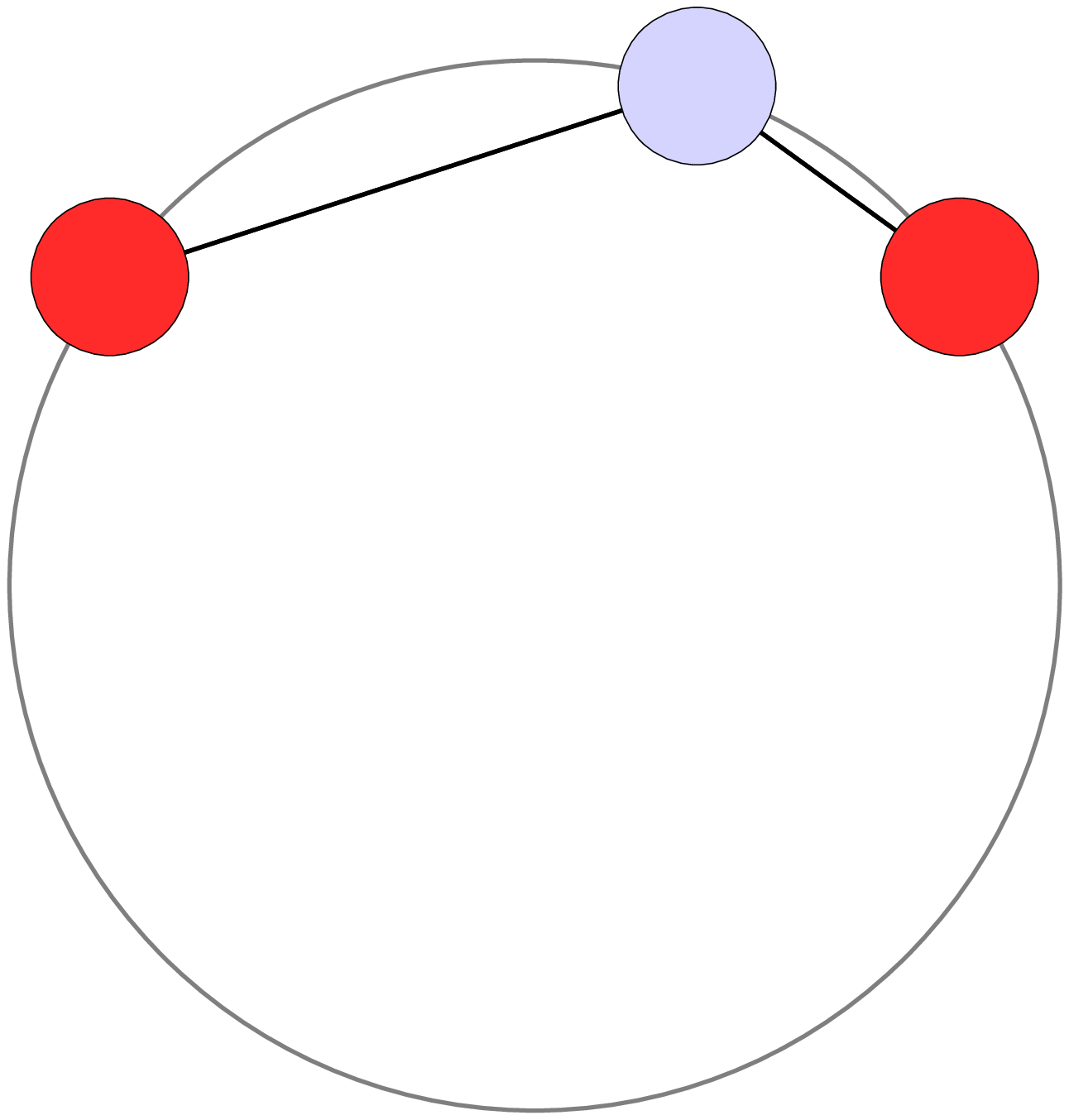}}
\caption{(Color online) A Type-I orbit (a) and a Type-II orbit (b) in 1d RGGs on the unit circle.
Vertices in an orbit are colored red.
Vertices in Type-I orbits are connected to each other, but vertices in Type-II orbits are not.}
\label{orbits}
\end{figure}

The network structure that causes a given eigenvalue can be identified by considering the 
corresponding eigenvector.  
In an RGG, the eigenvectors of many of the 
integer eigenvalues are exactly localized on symmetric motifs such that the eigenvector's only 
nonzero components correspond to vertices of the motif. 
Such motifs, where the adjacency of the vertices is 
invariant under a permutation of indices, are called graph orbits~\cite{redundancy}.  
Orbits leading to integer eigenvalues are significantly more frequent in spatial networks, such as RGGS, because of the geographical proximity of neighbors.  
The simplest type of orbit consists of a single set of vertices that can all be permuted.
If the vertices in such a simple orbit are all connected to each other, so that they 
form a clique,  
we call it a \emph{Type-I orbit}. If they are not connected to each other
then we call it a \emph{Type-II orbit}.
Simple examples of Type I and Type II orbits are shown in Fig.~\ref{orbits}.
More complicated types of orbits are also possible, but are rare in RGGs.

Type-I and Type-II orbits account for the vast majority 
of the integer eigenvalues found in 
the Laplacian spectra of RGGs.  
Since eigenvectors localize on the orbits, the corresponding 
eigenvalue is independent of the embedding network.  
Type-I orbits produce an eigenvalue equal to one 
more than the degree of the vertices in the orbit, while the eigenvalues resulting 
from Type-II orbits are equal to their degree~\cite{oldmath}.  
To see this, let $i$ and $j$ be two vertices in a simple orbit, and $\mathbf x$ be a 
vector with components $x_i=1$, $x_j=-1$, and all others zero.  If $k$ is 
the degree of the vertices in the orbit, then $\mathcal{L}\mathbf x =(k+1) \mathbf x$ 
for Type-I and $\mathcal{L}\mathbf x=k \mathbf x$ for Type-II.  
An orbit of size $n$ has $n-1$ of these independent and orthogonal eigenvectors.

To compute the expected number of simple orbits in 1d RGGs   
we begin by defining the following terms.
The \emph{geographical neighborhood} $\mathcal{N}(i)$ of a vertex $i$ is the 
region within a distance $r$ of the vertex.  
For a pair of vertices, their 
\emph{shared neighborhood} 
$\mathcal{N}_{\text{s}}(i,j)=\mathcal{N}(i) \cap \mathcal{N}(j)$ 
is the common region that is in the geographical neighborhood of 
both vertices and their 
\emph{excluded neighborhood} 
$\mathcal{N}_{\text{ex}}(i,j)
= (\mathcal{N}(i) \cup \mathcal{N}(j)) \setminus  
(\mathcal{N}(i) \cap \mathcal{N}(j))$ 
is the region that is in the 
geographical neighborhood of one vertex, but not the other.

The average multiplicity of eigenvalues due to Type-I orbits can be calculated by
considering each of the $N(N-1)/2$ pairs of vertices and calculating the 
likelihood that they are nearest neighbors with exactly $k-2$ vertices in their 
shared neighborhood and no vertices in their excluded neighborhood.  
Such a motif produces an integer eigenvalue equal to $k$.  
Requiring that the pair is a nearest neighbor pair ensures that the correct 
multiplicity of the eigenvalues is obtained because in a 1d RGG orbits that are 
chains of size $n$ have $n-1$ nearest neighbor pairs.    
Then, 
for $r \leq \frac13$,
the expected number of eigenvalues $\lambda=k$ is
\begin{eqnarray} \label{eq:typeiint}
E_1(k) 
& = & \frac{N!r(2r)^{k-2}(1-2r)^{N-k}}{(k-1)!(N-k)!} \nonumber \\
& & \qquad \times {}_2F_1\left[1,k-N,k,\tfrac{r}{1-2r}\right],
\end{eqnarray}
where ${}_2F_1$ is the ordinary hypergeometric function.
Because vertices are distributed randomly with uniform probability in RGGs, the probability
that a particular number of them are in a region of a given size is
given by the binomial distribution. 
Then, noting that Type-I orbits can produce eigenvalues ranging from 2 to $N$,
the total number of integer eigenvalues due to Type-I orbits is
\begin{eqnarray}
T_1 & = & \sum_{k=2}^N E_1(k) 
=\frac{N}{3} \left[1-\left(1-3r\right)^{N-1}\right] 
\label{eq:third}
\end{eqnarray}

In 1d RGGs, Type-II orbits and their shared neighbors always consist of an entire component of the graph.  
The average multiplicity of eigenvalues due to Type-II orbits can be calculated similarly
to those due to Type-I orbits, except that now one must calculate the probability
that a pair of vertices not connected to each other share exactly the same $k$ neighbors.  
For $r \leq \frac{1}{4}$,
the expected number of eigenvalues $\lambda=k$ is
\begin{eqnarray}\label{eq:type2smallr}
E_{2}(k)
&=&\frac{N!r^{k+1}(1-3r)^{N-2-k}}{(k+1)!(N-2-k)!} \nonumber \\
& & \qquad \times {}_2F_1\left[1,k+2-N,k+2,\tfrac{r}{1-3r}\right]. 
\end{eqnarray}
Type-II orbits can produce eigenvalues ranging from 0 to $N-2$,
thus
the total number of integer eigenvalues due to Type-II orbits is
\begin{eqnarray}
T_2& = &\sum_{k=0}^{N-2}E_{2}(k) 
=\frac{N}{2} \left[\left(1-2r\right)^{N-1}-(1-4r)^{N-1}\right]   
\label{t2}
\end{eqnarray}

Finally, in addition to the $\lambda=0$ eigenvalues due to Type-II orbits in which two vertices
are between $r$ and $2r$ apart and their shared neighborhood is empty,
there is an extra contribution of eigenvalue $\lambda=0$ in the spectra that 
occurs when pairs of vertices more than $2r$ apart 
have no edges at all.  
Such pairs are too far apart to have a shared neighborhood. 
The expected number of these extra eigenvalues is
\begin{equation}
E_*(0)=(N-1)(1-4r)^{N-1}.
\end{equation}
The contribution to the spectrum
is analogous to the accumulation of particles in the lowest 
energy level in Bose-Einstein condensation.  In the analogy, $r$
plays the role of temperature and the eigenvalues are the energy levels.  
Note that $\lim_{r \to 0}E_*(0)=N-1$.  

\begin{figure}
  \centering
  \includegraphics[trim = 0pt 0pt 0pt 0pt, clip=true, width=.4\textwidth]{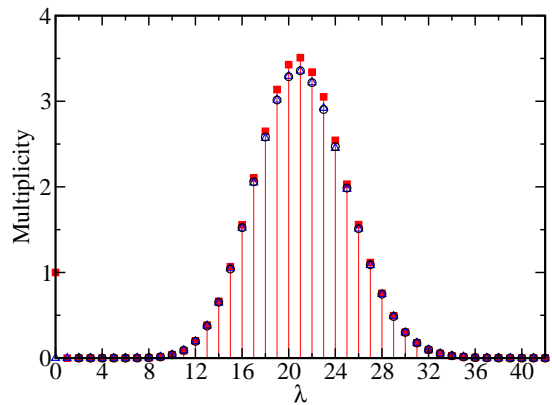}
  \caption{(Color online) Ensemble-averaged integer eigenvalue multiplicity for 1d RGGs with
  $N=100$ and $r=0.1$.  The red points are the results of numerical diagonalization 
of $10^6$ random ensemble realizations
  and the 
  black points are the number due to Type-I orbits according to Eq.~\ref{eq:typeiint}.} \label{fig:n100r1per}
\end{figure}

Figure~\ref{fig:n100r1per} shows a comparison of the average 
multiplicity of integer eigenvalues 
for an ensemble of 1d RGGs  
found using numerical diagonalization
with the number due to Type-I orbits predicted analytically using
Eq.~\ref{eq:typeiint}.
In the case considered, $N=100$ and $r=0.1$, 
the total number of eigenvalues due to Type-I orbits 
is $\sim 33.3$, which is $\sim 1/3$ of all eigenvalues.  
Note that for this case
the Type-I orbits account for the vast majority of integer eigenvalues,
as there are only $\sim 10^{-10}$ expected to be caused by Type-II orbits 
and only $\sim10^{-22}$ expected to be in the condensate of extra $\lambda=0$ 
eigenvalues.
The discrepancy between the theoretical and numerical curves 
in Fig.~\ref{fig:n100r1per}  arises 
due to the presence of orbits with more complex symmetries and 
other mechanisms~\cite{almostequitable}.

\begin{figure}
  \centering
  \includegraphics[trim = 0pt 0pt 0pt 0pt, clip=true, width=.4\textwidth]{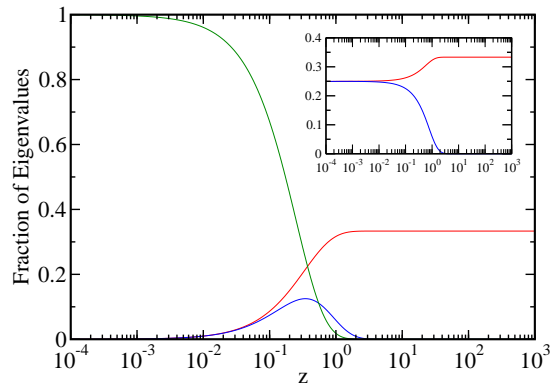}
  \caption{(Color online) Fraction of eigenvalues due to Type-I orbits (red), Type-II orbits (blue), and in 
  the $\lambda=0$ condensate (green) in the extensive large network limit as a function of 
  $z=Nr$. Inset shows the fraction of eigenvalues not in the $\lambda=0$ condensate that are
  due to Type-I (red) and Type-II (blue) orbits.} 
  \label{fig:al1}
\end{figure}
The fraction of the spectra due to simple orbits in large network limits
can be calculated using Eqs.~\ref{eq:third} and \ref{t2}. 
In the intensive limit of large $N$ and fixed $r$, 
for all $r$, 
$\lim_{N \to \infty}\frac{1}{N}T_2 = 0$ because there are no Type-II orbits,
but there are many Type-I orbits and the fraction of corresponding eigenvalues is
\begin{equation}
\lim_{N \to \infty}\frac{1}{N}T_1 = \frac13.  
\end{equation}
Thus, the discrete part of the spectrum comprises a substantial finite fraction 
of the total number of eigenvalues, even in the large network limit.

Perhaps a more important thermodynamic limit though is the extensive limit in which
 $N \rightarrow \infty$ while the average degree $Nr=z$ is constant. 
Figure~\ref{fig:al1} shows that, 
in this limit, 
for $z \gg 1$,
as in the nonextensive limit with fixed $r$, a third of the eigenvalues are due to
Type-I orbits, while virtually none are due to Type-II orbits and the $\lambda=0$ condensate
is empty. However, near the giant component transition, $z \sim 1$, the situation changes. 
Here the fraction of eigenvalues due
to Type-I starts to decrease, the $\lambda=0$ condensate starts to fill, and the 
fraction of eigenvalues due to Type-II orbits reaches a maximum. 
For $z \ll 1$, the condensate absorbs almost all of the eigenvalues, 
while the fraction of the eigenvalues that are due to simple orbits vanishes. 
However, as shown in the inset of Fig.~\ref{fig:al1}, even for $z \ll 1$, a substantial 
fraction of the eigenvalues that are not in the condensate are always 
eigenvalues due to simple orbits. In this limit 
Type-I and Type-II orbits each produce a quarter of the eigenvalues
that are not in the condensate.

In higher dimensional RGGs, similar calculations of the expected
number of eigenvalues due to simple orbits are not easy to do exactly.
However, what is generally true in higher dimensions
is that in the extensive thermodynamic limit
in which mean vertex degree is preserved 
the number of simple orbits and the integer eigenvalues they
produce are maximal when the mean degree is near 
the threshold of bond 
percolation where the giant component forms. 
This occurs when the mean degree is of order one.
In the limit of low mean degree almost all vertices are isolated 
and thus almost all eigenvalues are 
zero and part of the condensate.  As mean degree is increased, small clusters 
of vertices begin to form. Symmetries are plentiful in these small clusters. 
As the mean degree increases further, the number of components decreases with 
the giant component dominating the graph, and the number of both Type-I and Type-II 
orbits decrease.
In two-dimensional RGGs, we find numerically that the eigenvalues
due to Type-I orbits comprise about 12\% of the spectrum near the giant component transition
in connectivity.

The existence of such a large number of orbits has a profound effect on the way that
spatial networks behave dynamically. 
Laplacian spectra describe the normal modes of diffusion
on the network, and if the vertices are connected elastically, the normal modes of vibration.  
Generally the occurrence of symmetries in a networked system signals that, to linear order, its behavior can be decomposed into a contribution that is confined to a given symmetric structure, and a contribution which affects all vertices in an orbit equally. 
On the one hand, this impacts controllability of dynamical systems on spatial graphs as the confined modes cannot be excited to linear order by a controller situated elsewhere~\cite{control}. 
On the other hand, the confinement of dynamical modes inside the symmetric structure means that mesoscale structures can exist whose dynamics are largely independent of the embedding network, but which may be able to communicate with the embedding network through nonlinear effects. 
One can imagine that thereby reusable structures are created that can perform the same function independently of the surrounding networks. 
Symmetric structures in dynamical networks could therefore have an important effect, for instance, on the evolvability of biological systems. 
They also have important consequences, for example, in 
neural networks~\cite{toro1,toro2}, where an orbit implies that that neurons can be 
locally excited, and wireless communication networks~\cite{rgg_sync,rgg_mobile}.

The existence of such a large number of orbits in spatial networks can also be used
to simplify the analysis of their behavior. By considering a quotient graph~\cite{redundancy,lars}
the integer eigenvalues can be removed, leaving only the continuous part of the
spectrum. The continuous part of the spectrum describes much of the important properties
of the network's behavior. For example, the smallest nonzero eigenvalue in graph
Laplacian matrices determines the number of vertices or edges that must be cut to
sever the network~\cite{fiedler}. Also, its eigenvector can be used to partition the network into 
communities~\cite{partition}. Notice from Fig.~2 that, in the case shown, pairs of eigenvalues
split-off or separate~\cite{BFF09,BFF10} from the bulk continuous distribution at the 
small end of the spectrum. These separated eigenvalues include the smallest nonzero one.
 The number of
eigenvalues that split off together from the bulk continuous distribution can be
deduced by approximating the graph Laplacian with a continuous Laplacian operator 
corresponding to disordered random media and considering the degeneracy of the modes
with smallest eigenvalues~\cite{NBupub}.

Thus, we have shown analytically that symmetric mesoscale structures are highly
abundant for all $r$ in 1d RGGs. These motifs lead to integer eigenvalues that comprise
a substantial fraction, more than a third, 
in both intensive and extensive large network limits.
Approximate arguments and 
numerical results indicate that similar behavior occurs in higher dimensional RGGs as well.
This behavior differs remarkably from that of non-spatial graphs, which in thermodynamic
limits have almost no orbits.

The work of AN and KEB was supported by the NSF through grant DMR-1206839, and by
the AFSOR and DARPA through grant FA9550-12-1-0405.  

\bibliography{intbib}

\begin{thebibliography}{34}
\expandafter\ifx\csname natexlab\endcsname\relax\def\natexlab#1{#1}\fi
\expandafter\ifx\csname bibnamefont\endcsname\relax
  \def\bibnamefont#1{#1}\fi
\expandafter\ifx\csname bibfnamefont\endcsname\relax
  \def\bibfnamefont#1{#1}\fi
\expandafter\ifx\csname citenamefont\endcsname\relax
  \def\citenamefont#1{#1}\fi
\expandafter\ifx\csname url\endcsname\relax
  \def\url#1{\texttt{#1}}\fi
\expandafter\ifx\csname urlprefix\endcsname\relax\def\urlprefix{URL }\fi
\providecommand{\bibinfo}[2]{#2}
\providecommand{\eprint}[2][]{\url{#2}}

\bibitem[{\citenamefont{Albert and Barab\'asi}(2002)}]{albar}
\bibinfo{author}{\bibfnamefont{R.}~\bibnamefont{Albert}} \bibnamefont{and}
  \bibinfo{author}{\bibfnamefont{A.-L.} \bibnamefont{Barab\'asi}},
  \bibinfo{journal}{Rev. Mod. Phys.} \textbf{\bibinfo{volume}{74}},
  \bibinfo{pages}{47} (\bibinfo{year}{2002}).

\bibitem[{\citenamefont{Newman}(2003)}]{newmanrev}
\bibinfo{author}{\bibfnamefont{M.~E.~J.} \bibnamefont{Newman}},
  \bibinfo{journal}{SIAM Review} \textbf{\bibinfo{volume}{45}},
  \bibinfo{pages}{pp. 167} (\bibinfo{year}{2003}).

\bibitem[{\citenamefont{Boccaletti et~al.}(2006)\citenamefont{Boccaletti,
  Latora, Moreno, Chavez, and Hwang}}]{boccaletti}
\bibinfo{author}{\bibfnamefont{S.}~\bibnamefont{Boccaletti}},
  \bibinfo{author}{\bibfnamefont{V.}~\bibnamefont{Latora}},
  \bibinfo{author}{\bibfnamefont{Y.}~\bibnamefont{Moreno}},
  \bibinfo{author}{\bibfnamefont{M.}~\bibnamefont{Chavez}}, \bibnamefont{and}
  \bibinfo{author}{\bibfnamefont{D.-U.} \bibnamefont{Hwang}},
  \bibinfo{journal}{Physics Reports} \textbf{\bibinfo{volume}{424}},
  \bibinfo{pages}{175 } (\bibinfo{year}{2006}).

\bibitem[{\citenamefont{Almendral et~al.}(2011)\citenamefont{Almendral, Criado,
  Leyva, Buld\'u, and Sendi\~na Nadal}}]{chaosrev}
\bibinfo{author}{\bibfnamefont{J.~A.} \bibnamefont{Almendral}},
  \bibinfo{author}{\bibfnamefont{R.}~\bibnamefont{Criado}},
  \bibinfo{author}{\bibfnamefont{I.}~\bibnamefont{Leyva}},
  \bibinfo{author}{\bibfnamefont{J.~M.} \bibnamefont{Buld\'u}},
  \bibnamefont{and} \bibinfo{author}{\bibfnamefont{I.}~\bibnamefont{Sendi\~na
  Nadal}}, \bibinfo{journal}{Chaos} \textbf{\bibinfo{volume}{21}},
  \bibinfo{pages}{016101} (\bibinfo{year}{2011}).

\bibitem[{\citenamefont{Zanin et~al.}(2014)\citenamefont{Zanin, Sousa, and
  Menasalvas}}]{ic}
\bibinfo{author}{\bibfnamefont{M.}~\bibnamefont{Zanin}},
  \bibinfo{author}{\bibfnamefont{P.~A.} \bibnamefont{Sousa}}, \bibnamefont{and}
  \bibinfo{author}{\bibfnamefont{E.}~\bibnamefont{Menasalvas}},
  \bibinfo{journal}{EPL} \textbf{\bibinfo{volume}{106}}, \bibinfo{pages}{30001}
  (\bibinfo{year}{2014}).

\bibitem[{\citenamefont{Chung}(1997)}]{chung}
\bibinfo{author}{\bibfnamefont{F.}~\bibnamefont{Chung}},
  \emph{\bibinfo{title}{Spectral Graph Theory}}, vol.~\bibinfo{volume}{92} of
  \emph{\bibinfo{series}{CBMS Regional conference series in mathematics}}
  (\bibinfo{publisher}{Conference Board of the Mathematical Sciences},
  \bibinfo{address}{Washington, DC}, \bibinfo{year}{1997}).

\bibitem[{\citenamefont{Farkas et~al.}(2001)\citenamefont{Farkas, Der\'enyi,
  Barab\'asi, and Vicsek}}]{beyond}
\bibinfo{author}{\bibfnamefont{I.~J.} \bibnamefont{Farkas}},
  \bibinfo{author}{\bibfnamefont{I.}~\bibnamefont{Der\'enyi}},
  \bibinfo{author}{\bibfnamefont{A.-L.} \bibnamefont{Barab\'asi}},
  \bibnamefont{and} \bibinfo{author}{\bibfnamefont{T.}~\bibnamefont{Vicsek}},
  \bibinfo{journal}{Phys. Rev. E} \textbf{\bibinfo{volume}{64}},
  \bibinfo{pages}{026704} (\bibinfo{year}{2001}).

\bibitem[{\citenamefont{Pecora and Carroll}(1998)}]{pecora}
\bibinfo{author}{\bibfnamefont{L.~M.} \bibnamefont{Pecora}} \bibnamefont{and}
  \bibinfo{author}{\bibfnamefont{T.~L.} \bibnamefont{Carroll}},
  \bibinfo{journal}{Phys. Rev. Lett.} \textbf{\bibinfo{volume}{80}},
  \bibinfo{pages}{2109} (\bibinfo{year}{1998}).

\bibitem[{\citenamefont{Mieghem}(2011)}]{van}
\bibinfo{author}{\bibfnamefont{P.}~\bibnamefont{Mieghem}},
  \emph{\bibinfo{title}{Graph Spectra for Complex Networks}}
  (\bibinfo{publisher}{Cambridge University Press}, \bibinfo{year}{2011}).

\bibitem[{\citenamefont{MacArthur and
  S\'anchez-Garc\'\i{}a}(2009)}]{redundancy}
\bibinfo{author}{\bibfnamefont{B.~D.} \bibnamefont{MacArthur}}
  \bibnamefont{and} \bibinfo{author}{\bibfnamefont{R.~J.}
  \bibnamefont{S\'anchez-Garc\'\i{}a}}, \bibinfo{journal}{Phys. Rev. E}
  \textbf{\bibinfo{volume}{80}}, \bibinfo{pages}{026117}
  (\bibinfo{year}{2009}).

\bibitem[{\citenamefont{MacArthur et~al.}(2008)\citenamefont{MacArthur,
  S\'anchez-Garc\'ia, and Anderson}}]{symmetry}
\bibinfo{author}{\bibfnamefont{B.~D.} \bibnamefont{MacArthur}},
  \bibinfo{author}{\bibfnamefont{R.~J.} \bibnamefont{S\'anchez-Garc\'ia}},
  \bibnamefont{and} \bibinfo{author}{\bibfnamefont{J.~W.}
  \bibnamefont{Anderson}}, \bibinfo{journal}{Discrete Applied Mathematics}
  \textbf{\bibinfo{volume}{156}}, \bibinfo{pages}{3525 }
  (\bibinfo{year}{2008}).

\bibitem[{\citenamefont{Arenas et~al.}(2006)\citenamefont{Arenas,
  D\'iaz-Guilera, and P\'erez-Vicente}}]{arenas}
\bibinfo{author}{\bibfnamefont{A.}~\bibnamefont{Arenas}},
  \bibinfo{author}{\bibfnamefont{A.}~\bibnamefont{D\'iaz-Guilera}},
  \bibnamefont{and} \bibinfo{author}{\bibfnamefont{C.~J.}
  \bibnamefont{P\'erez-Vicente}}, \bibinfo{journal}{Physica D: Nonlinear
  Phenomena} \textbf{\bibinfo{volume}{224}}, \bibinfo{pages}{27 }
  (\bibinfo{year}{2006}).

\bibitem[{\citenamefont{D\'{i}az-Guilera
  et~al.}(2009)\citenamefont{D\'{i}az-Guilera, G\'omez-Garde\~nes, Moreno, and
  Nekovee}}]{rgg_sync}
\bibinfo{author}{\bibfnamefont{A.}~\bibnamefont{D\'{i}az-Guilera}},
  \bibinfo{author}{\bibfnamefont{J.}~\bibnamefont{G\'omez-Garde\~nes}},
  \bibinfo{author}{\bibfnamefont{Y.}~\bibnamefont{Moreno}}, \bibnamefont{and}
  \bibinfo{author}{\bibfnamefont{M.}~\bibnamefont{Nekovee}},
  \bibinfo{journal}{International Journal of Bifurcation and Chaos}
  \textbf{\bibinfo{volume}{19}}, \bibinfo{pages}{687} (\bibinfo{year}{2009}).

\bibitem[{\citenamefont{Aufderheide et~al.}(2012)\citenamefont{Aufderheide,
  Rudolf, and Gross}}]{lars}
\bibinfo{author}{\bibfnamefont{H.}~\bibnamefont{Aufderheide}},
  \bibinfo{author}{\bibfnamefont{L.}~\bibnamefont{Rudolf}}, \bibnamefont{and}
  \bibinfo{author}{\bibfnamefont{T.}~\bibnamefont{Gross}},
  \bibinfo{journal}{New Journal of Physics} \textbf{\bibinfo{volume}{14}},
  \bibinfo{pages}{105014} (\bibinfo{year}{2012}).

\bibitem[{\citenamefont{Do et~al.}(2012)\citenamefont{Do, H\"{o}fener, and
  Gross}}]{Ly}
\bibinfo{author}{\bibfnamefont{A.-L.} \bibnamefont{Do}},
  \bibinfo{author}{\bibfnamefont{J.}~\bibnamefont{H\"{o}fener}},
  \bibnamefont{and} \bibinfo{author}{\bibfnamefont{T.}~\bibnamefont{Gross}},
  \bibinfo{journal}{New Journal of Physics} \textbf{\bibinfo{volume}{14}},
  \bibinfo{pages}{115022} (\bibinfo{year}{2012}).

\bibitem[{\citenamefont{Dall and Christensen}(2002)}]{rgg_dall}
\bibinfo{author}{\bibfnamefont{J.}~\bibnamefont{Dall}} \bibnamefont{and}
  \bibinfo{author}{\bibfnamefont{M.}~\bibnamefont{Christensen}},
  \bibinfo{journal}{Phys. Rev. E} \textbf{\bibinfo{volume}{66}},
  \bibinfo{pages}{016121} (\bibinfo{year}{2002}).

\bibitem[{\citenamefont{Penrose}(2003)}]{penrose}
\bibinfo{author}{\bibfnamefont{M.}~\bibnamefont{Penrose}},
  \emph{\bibinfo{title}{Random geometric graphs}}, Oxford studies in
  probability (\bibinfo{publisher}{Oxford University Press},
  \bibinfo{year}{2003}).

\bibitem[{\citenamefont{Barth\'elemy}(2011)}]{spatial1}
\bibinfo{author}{\bibfnamefont{M.}~\bibnamefont{Barth\'elemy}},
  \bibinfo{journal}{Phys. Rep.} \textbf{\bibinfo{volume}{499}},
  \bibinfo{pages}{1} (\bibinfo{year}{2011}).

\bibitem[{\citenamefont{Bullock et~al.}(2010)\citenamefont{Bullock, Barnett,
  and Di~Paolo}}]{spatial2}
\bibinfo{author}{\bibfnamefont{S.}~\bibnamefont{Bullock}},
  \bibinfo{author}{\bibfnamefont{L.}~\bibnamefont{Barnett}}, \bibnamefont{and}
  \bibinfo{author}{\bibfnamefont{E.~A.} \bibnamefont{Di~Paolo}},
  \bibinfo{journal}{Complexity} \textbf{\bibinfo{volume}{16}},
  \bibinfo{pages}{20} (\bibinfo{year}{2010}).

\bibitem[{\citenamefont{Haenggi et~al.}(2009)\citenamefont{Haenggi, Andrews,
  Baccelli, Dousse, and Franceschetti}}]{wireless}
\bibinfo{author}{\bibfnamefont{M.}~\bibnamefont{Haenggi}},
  \bibinfo{author}{\bibfnamefont{J.}~\bibnamefont{Andrews}},
  \bibinfo{author}{\bibfnamefont{F.}~\bibnamefont{Baccelli}},
  \bibinfo{author}{\bibfnamefont{O.}~\bibnamefont{Dousse}}, \bibnamefont{and}
  \bibinfo{author}{\bibfnamefont{M.}~\bibnamefont{Franceschetti}},
  \bibinfo{journal}{IEEE Journal on Selected Areas in Communications}
  \textbf{\bibinfo{volume}{27}}, \bibinfo{pages}{1029 } (\bibinfo{year}{2009}).

\bibitem[{\citenamefont{Xiao and Yeh}(2011)}]{grids}
\bibinfo{author}{\bibfnamefont{H.}~\bibnamefont{Xiao}} \bibnamefont{and}
  \bibinfo{author}{\bibfnamefont{E.}~\bibnamefont{Yeh}},
  \bibinfo{journal}{Communications Workshops (ICC), 2011 IEEE International
  Conference on} pp. \bibinfo{pages}{1--6} (\bibinfo{year}{2011}).

\bibitem[{\citenamefont{Markov et~al.}(2013)\citenamefont{Markov,
  Ercsey-Ravasz, Van~Essen, Knoblauch, Toroczkai, and Kennedy}}]{toro1}
\bibinfo{author}{\bibfnamefont{N.~T.} \bibnamefont{Markov}},
  \bibinfo{author}{\bibfnamefont{M.}~\bibnamefont{Ercsey-Ravasz}},
  \bibinfo{author}{\bibfnamefont{D.~C.} \bibnamefont{Van~Essen}},
  \bibinfo{author}{\bibfnamefont{K.}~\bibnamefont{Knoblauch}},
  \bibinfo{author}{\bibfnamefont{Z.}~\bibnamefont{Toroczkai}},
  \bibnamefont{and} \bibinfo{author}{\bibfnamefont{H.}~\bibnamefont{Kennedy}},
  \bibinfo{journal}{Science} \textbf{\bibinfo{volume}{342}}
  (\bibinfo{year}{2013}).

\bibitem[{\citenamefont{Ercsey-Ravasz et~al.}(2013)\citenamefont{Ercsey-Ravasz,
  Markov, Lamy, Essen, Knoblauch, Toroczkai, and Kennedy}}]{toro2}
\bibinfo{author}{\bibfnamefont{M.}~\bibnamefont{Ercsey-Ravasz}},
  \bibinfo{author}{\bibfnamefont{N.~T.} \bibnamefont{Markov}},
  \bibinfo{author}{\bibfnamefont{C.}~\bibnamefont{Lamy}},
  \bibinfo{author}{\bibfnamefont{D.~C.~V.} \bibnamefont{Essen}},
  \bibinfo{author}{\bibfnamefont{K.}~\bibnamefont{Knoblauch}},
  \bibinfo{author}{\bibfnamefont{Z.}~\bibnamefont{Toroczkai}},
  \bibnamefont{and} \bibinfo{author}{\bibfnamefont{H.}~\bibnamefont{Kennedy}},
  \bibinfo{journal}{Neuron} \textbf{\bibinfo{volume}{80}}, \bibinfo{pages}{184
  } (\bibinfo{year}{2013}), ISSN \bibinfo{issn}{0896-6273}.

\bibitem[{\citenamefont{Higham et~al.}(2008)\citenamefont{Higham, Ra\v{s}ajski,
  and Pr\v{z}ulj}}]{bio1}
\bibinfo{author}{\bibfnamefont{D.~J.} \bibnamefont{Higham}},
  \bibinfo{author}{\bibfnamefont{M.}~\bibnamefont{Ra\v{s}ajski}},
  \bibnamefont{and}
  \bibinfo{author}{\bibfnamefont{N.}~\bibnamefont{Pr\v{z}ulj}},
  \bibinfo{journal}{Bioinformatics} \textbf{\bibinfo{volume}{24}},
  \bibinfo{pages}{1093} (\bibinfo{year}{2008}).

\bibitem[{\citenamefont{Aguirre et~al.}(2011)\citenamefont{Aguirre, Buld\'u,
  Stich, and Manrubia}}]{bio2}
\bibinfo{author}{\bibfnamefont{J.}~\bibnamefont{Aguirre}},
  \bibinfo{author}{\bibfnamefont{J.~M.} \bibnamefont{Buld\'u}},
  \bibinfo{author}{\bibfnamefont{M.}~\bibnamefont{Stich}}, \bibnamefont{and}
  \bibinfo{author}{\bibfnamefont{S.~C.} \bibnamefont{Manrubia}},
  \bibinfo{journal}{PLoS ONE} \textbf{\bibinfo{volume}{6}},
  \bibinfo{pages}{e26324} (\bibinfo{year}{2011}).

\bibitem[{\citenamefont{Grone and Merris}(1994)}]{oldmath}
\bibinfo{author}{\bibfnamefont{R.}~\bibnamefont{Grone}} \bibnamefont{and}
  \bibinfo{author}{\bibfnamefont{R.}~\bibnamefont{Merris}},
  \bibinfo{journal}{SIAM J. Discrete Math.} \textbf{\bibinfo{volume}{7}},
  \bibinfo{pages}{221} (\bibinfo{year}{1994}).

\bibitem[{\citenamefont{Cardoso et~al.}(2007)\citenamefont{Cardoso, Delorme,
  and Rama}}]{almostequitable}
\bibinfo{author}{\bibfnamefont{D.~M.} \bibnamefont{Cardoso}},
  \bibinfo{author}{\bibfnamefont{C.}~\bibnamefont{Delorme}}, \bibnamefont{and}
  \bibinfo{author}{\bibfnamefont{P.}~\bibnamefont{Rama}},
  \bibinfo{journal}{European Journal of Combinatorics}
  \textbf{\bibinfo{volume}{28}}, \bibinfo{pages}{665 } (\bibinfo{year}{2007}).

\bibitem[{\citenamefont{Liu et~al.}(2011)\citenamefont{Liu, Slotine, and
  Barab\'asi}}]{control}
\bibinfo{author}{\bibfnamefont{Y.-Y.} \bibnamefont{Liu}},
  \bibinfo{author}{\bibfnamefont{J.-J.} \bibnamefont{Slotine}},
  \bibnamefont{and} \bibinfo{author}{\bibfnamefont{A.-L.}
  \bibnamefont{Barab\'asi}}, \bibinfo{journal}{Nature}
  \textbf{\bibinfo{volume}{473}}, \bibinfo{pages}{167} (\bibinfo{year}{2011}).

\bibitem[{\citenamefont{Fujiwara et~al.}(2011)\citenamefont{Fujiwara, Kurths,
  and D\'\i{}az-Guilera}}]{rgg_mobile}
\bibinfo{author}{\bibfnamefont{N.}~\bibnamefont{Fujiwara}},
  \bibinfo{author}{\bibfnamefont{J.}~\bibnamefont{Kurths}}, \bibnamefont{and}
  \bibinfo{author}{\bibfnamefont{A.}~\bibnamefont{D\'\i{}az-Guilera}},
  \bibinfo{journal}{Phys. Rev. E} \textbf{\bibinfo{volume}{83}},
  \bibinfo{pages}{025101} (\bibinfo{year}{2011}).

\bibitem[{\citenamefont{Fiedler}(1973)}]{fiedler}
\bibinfo{author}{\bibfnamefont{M.}~\bibnamefont{Fiedler}},
  \bibinfo{journal}{Czechoslovak Mathematical Journal}
  \textbf{\bibinfo{volume}{23}}, \bibinfo{pages}{298} (\bibinfo{year}{1973}).

\bibitem[{\citenamefont{Spielman and Teng}(2007)}]{partition}
\bibinfo{author}{\bibfnamefont{D.~A.} \bibnamefont{Spielman}} \bibnamefont{and}
  \bibinfo{author}{\bibfnamefont{S.-H.} \bibnamefont{Teng}},
  \bibinfo{journal}{Linear Algebra and its Applications}
  \textbf{\bibinfo{volume}{421}}, \bibinfo{pages}{284 } (\bibinfo{year}{2007}).

\bibitem[{\citenamefont{Bassler et~al.}(2009)\citenamefont{Bassler, Forrester,
  and Frankel}}]{BFF09}
\bibinfo{author}{\bibfnamefont{K.~E.} \bibnamefont{Bassler}},
  \bibinfo{author}{\bibfnamefont{P.~J.} \bibnamefont{Forrester}},
  \bibnamefont{and} \bibinfo{author}{\bibfnamefont{N.~E.}
  \bibnamefont{Frankel}}, \bibinfo{journal}{Journal of Mathematical Physics}
  \textbf{\bibinfo{volume}{50}}, \bibinfo{eid}{033302} (\bibinfo{year}{2009}).

\bibitem[{\citenamefont{Bassler et~al.}(2010)\citenamefont{Bassler, Forrester,
  and Frankel}}]{BFF10}
\bibinfo{author}{\bibfnamefont{K.~E.} \bibnamefont{Bassler}},
  \bibinfo{author}{\bibfnamefont{P.~J.} \bibnamefont{Forrester}},
  \bibnamefont{and} \bibinfo{author}{\bibfnamefont{N.~E.}
  \bibnamefont{Frankel}}, \bibinfo{journal}{Journal of Mathematical Physics}
  \textbf{\bibinfo{volume}{51}}, \bibinfo{eid}{123305} (\bibinfo{year}{2010}).

\bibitem[{\citenamefont{Nyberg and Bassler}(2014)}]{NBupub}
\bibinfo{author}{\bibfnamefont{A.}~\bibnamefont{Nyberg}} \bibnamefont{and}
  \bibinfo{author}{\bibfnamefont{K.~E.} \bibnamefont{Bassler}},
  \bibinfo{journal}{unpublished}  (\bibinfo{year}{2014}).

\end{thebibliography}

\end{document}